\documentstyle[twocolumn,aps,prl]{revtex}

\begin{document}

\tighten
\draft

\title{Theory of conductance and noise additivity in parallel
       mesoscopic conductors}
\author{G. Iannaccone, M. Macucci, B. Pellegrini}

\address{
Dipartimento di Ingegneria dell'Informazione: 
Elettronica, Informatica e Telecomunicazioni \\
Universit\`a degli studi di Pisa, Via Diotisalvi 2, 
I-56126 Pisa, Italy}
\date{\today}
\maketitle

\begin{abstract} 
We present a theory of conductance and noise
in generic mesoscopic conductors connected in parallel,
and we demonstrate that the additivity of conductance 
and of shot noise arises as a sole property of the
junctions connecting the two (or more) conductors in parallel.
Consequences on the functionality of devices based on
the Aharonov-Bohm effect are also drawn.
\end{abstract}



Electron transport is said to be in the ballistic regime when
the phase coherence of the wave functions is maintained throughout
the whole extension of the structure, in other words, when the 
inelastic mean free path is greater than the size of the device.

Recent advances in nanofabrication techniques, and the possibility
of reaching operating temperatures in the millikelvin range (in order
to suppress phonon scattering),
have made feasible the study of transport in the ballistic regime,
and the exploration of the novel and interesting phenomena which emerge
when such regime is approached.

Since the pioneering work of van Wees and co-workers 
on conductance quantization \cite{vanwvanh88}, this transport regime
has been the subject of widespread interest, both from a theoretical and
an experimental point of view.

In this Letter, we focus on transport and noise in 
mesoscopic conductors connected in parallel. It suffices to point
out that such topology applies to all devices based on some kind 
of Aharonov-Bohm effect.

In the case of macroscopic conductors connected in parallel it is
well known that both the conductances and the shot-noise 
current power spectral densities add. 

While transport properties of macroscopic conductors depend on
local material properties, those of mesoscopic structures are  
obtained as the solution of a complex scattering problem, 
in which the shape of the boundaries and the potential profile over 
the whole device region play a relevant role.
Therefore, the problem needs to be reformulated in these new terms.
 
Numerical studies of conductance additivity in sample structures
made of two parallel constrictions,
along with some analytical justifications, exist in the literature
\cite{castkirc90,zhenberg91,macuhess92}.
Furthermore, a numerical study has been presented showing additivity of 
shot noise in parallel constrictions \cite{macucci96}.

We present a theory of conductance and noise
in generic mesoscopic conductors connected in parallel,
and we demonstrate that the additivity of conductance 
and of shot noise arises as a sole property of the
junctions connecting the two (or more) conductors in parallel.

In particular, additivity requires that 
the scattering matrix of both junctions (each of which is
a $(n+1)$-lead mesoscopic system, if the whole device is made up
of $n$ conductors in parallel), is such that zero conductance appears 
between the leads connecting different conductors in parallel. 
In such a case, even if phase coherence is maintained throughout the whole
device, there are no quantum-interference interactions between
parallel conductors.

With the purpose of showing that this requirement is not hard to meet,
we end the paper including the example of a junction 
warranting additivity. We also point out the
consequences of our results on evaluating the functionality and 
the noise properties of devices based on the Aharonov-Bohm effect.


The structure considered is sketched in Fig. 1. It consists of 2
mesoscopic conductors $\Sigma^u$ and $\Sigma^d$ connected in parallel
by means of two junctions, $\Sigma^l$ and $\Sigma^r$. The junctions
are ballistic systems with three leads, one connected to $\Sigma^u$,
one to $\Sigma^d$, and the other used as an external lead of the 
whole structure. Phase coherence is mantained in the whole system.

The internal and external leads of the whole structure are numbered from
1 to 6, as sketched in Fig. 1. Let us define
${\bf a_i}$ and ${\bf b_i}$ ($i=1,\dots,6$)
as the column vectors whose $N_i$ elements are the amplitudes of the modes
in lead $i$ entering and exiting the adjacent junction, respectively.

Electron transport in each of the subsystems $\Sigma^\alpha$ ($\alpha=u,d,l,r$)
is completely described by the associated scattering matrix $S^\alpha$,
\cite{davydov} which is unitary and such as $S^{\alpha T}({\bf B}) =
S^{\alpha}(-{\bf B})$, where ${\bf B}$ is the applied magnetic
field \cite{buettiker86}.

Let us first consider the case in which we have only the left junction
$\Sigma^l$ and leads 1, 2 and 3 are connected to different electron
reservoirs. The relation between incoming and outgoing modes can be
written as 
\begin{equation}
\left[ \begin{array}{c}
       {\bf b_1} \\
       {\bf b_2} \\
       {\bf b_3} 
       \end{array}
\right]= 
S^l 
\left[ \begin{array}{c}
       {\bf a_1} \\
       {\bf a_2} \\
       {\bf a_3} 
       \end{array}
\right] 
\end{equation}
if $S^l$ is arranged in the following way
\begin{equation}
S_l = \left[
      \begin{array}{ccc}
      s^l_{11} & s^l_{12} & s^l_{13} \\
      s^l_{21} & s^l_{22} & s^l_{23} \\
      s^l_{31} & s^l_{32} & s^l_{33} 
      \end{array}
      \right]
,\end{equation}
where $s^l_{ij}$ $(i,j=1,2,3)$ is a $N_i \times N_j$ matrix, relating the
amplitudes of the outgoing modes in lead $i$ to the amplitudes
of the incoming modes in lead $j$ (as many evanescent modes as needed
may be considered).

We can repeat the same considerations for the right junction: if leads
4, 5 and 6 are connected to different electron reservoirs we can write
\begin{equation}
\left[ \begin{array}{c}
       {\bf b_4} \\
       {\bf b_5} \\
       {\bf b_6} 
       \end{array}
\right]= 
\left[ \begin{array}{ccc}
       s^r_{44} & s^r_{45} & s^r_{46} \\
       s^r_{54} & s^r_{55} & s^r_{56} \\
       s^r_{64} & s^r_{65} & s^r_{66} 
       \end{array}
\right]
\left[ \begin{array}{c}
       {\bf a_4} \\
       {\bf a_5} \\
       {\bf a_6} 
       \end{array}
\right]
,\end{equation}
where we have already written $S^r$ in the form of 
submatrices $s^r_{ij}$, $(i,j = 4,5,6)$.
Analogously, for the conductors $\Sigma^u$ and $\Sigma^d$ we can write
\begin{equation}
\left[ \begin{array}{c}
       {\bf a_2} \\
       {\bf a_4} \\ 
       \end{array}
\right]= 
S^u 
\left[ \begin{array}{c}
       {\bf b_2} \\
       {\bf b_4} \\
       \end{array}
\right]
=
\left[ \begin{array}{cc}
       s^u_{22} & s^u_{24} \\
       s^u_{42} & s^u_{44} 
       \end{array}
\right]
\left[ \begin{array}{c}
       {\bf b_2} \\
       {\bf b_4} \\
       \end{array}
\right]
,\end{equation}
\begin{equation}
\left[ \begin{array}{c}
       {\bf a_3} \\
       {\bf a_5} \\ 
       \end{array}
\right]= 
S^d 
\left[ \begin{array}{c}
       {\bf b_3} \\
       {\bf b_5} \\
       \end{array}
\right]
=
\left[ \begin{array}{cc}
       s^d_{33} & s^d_{35} \\
       s^d_{53} & s^d_{55} 
       \end{array}
\right]
\left[ \begin{array}{c}
       {\bf b_3} \\
       {\bf b_5} \\
       \end{array}
\right]
.\end{equation}


We intend to demonstrate that conductances and noise 
in parallel conductors add if the scattering matrices of the 
junctions satisfy the following conditions:
\begin{eqnarray}
s^l_{32} = 0, & \hspace{1cm} & s^r_{45} = 0, \nonumber \\ 
s^l_{23} = 0, & \hspace{1cm} & s^r_{54} = 0 
.\label{condizioni}
\end{eqnarray}

Let us point out that in the absence of magnetic field 
the last two conditions are redundant.
The physical meaning of (\ref{condizioni}) is that an 
electron injected
in $\Sigma^l$ from lead 3 does not exit from lead 2 (and vice-versa),
and that an electron injected in $\Sigma^r$ from lead 5 is not
transmitted to lead 4 (and vice-versa).

Since $S^{l\dagger} S^l = 1$ and $S^{r\dagger} S^r = 1$ (i.e., the $S$-matrices
are unitary), by multiplying row by column 
and using (\ref{condizioni}), we straightforwardly obtain 
\begin{eqnarray}
s^{l\dagger}_{21} s^l_{31} = 0, 
& \hspace{1cm} &  s^l_{21} s^{l\dagger}_{31} = 0, \label{propleft} \\
s^{r\dagger}_{64} s^r_{65} = 0, 
& \hspace{1cm} &  s^r_{64} s^{r\dagger}_{65} = 0. \label{propright}  
\end{eqnarray}

In order to proceed with our demonstration we have to consider the
whole structure, and the associated scattering matrix 
$S_{\rm tot}$. We can write
\begin{equation}
\left[ \begin{array}{c}
       {\bf b_1} \\
       {\bf b_6} \\ 
       \end{array}
\right]= 
S_{\rm tot} 
\left[ \begin{array}{c}
       {\bf a_1} \\
       {\bf a_6} \\
       \end{array}
\right]
=
\left[ \begin{array}{cc}
       s_{11} & s_{16} \\
       s_{61} & s_{66} 
       \end{array}
\right]
\left[ \begin{array}{c}
       {\bf a_1} \\
       {\bf a_6} \\
       \end{array}
\right]
.\end{equation}

In addition, we have to consider the scattering matrices $S_{\rm up}$
and $S_{\rm down}$, corresponding to the whole system
with the conductor $\Sigma^d$ or $\Sigma^u$ removed, respectively.
They are of the form
\begin{equation}
S_{\rm up}
=
\left[ \begin{array}{cc}
       s^u_{11} & s^u_{16} \\
       s^u_{61} & s^u_{66} 
       \end{array}
\right]
, \hspace{1cm}
S_{\rm down}
=
\left[ \begin{array}{cc}
       s^d_{11} & s^d_{16} \\
       s^d_{61} & s^d_{66} 
       \end{array}
\right]
.\end{equation}

Let us start by calculating $s^u_{61}$, which relates the amplitudes
of the modes exiting from lead 6 to those of the modes entering from
lead 1: since $\Sigma^d$ has been removed
the only path from lead 1 to lead 6 is that through the upper
conductor; moreover, the hypothesis (\ref{condizioni}) implies
that the terminations of leads 3 and 5 have no effect on the
calculation of  $s^u_{61}$.
It is straighforward to write $s^u_{61}$ in the form
of the scattering series
\begin{eqnarray}
s^u_{61} & = &  
s^r_{64} s^u_{42} s^l_{21}
+ s^r_{64} s^u_{42} s^l_{22} s^u_{24} s^r_{44} s^u_{42} s^l_{21} +
\dots + \nonumber \\
& & s^r_{64} s^u_{42} (s^l_{22} s^u_{24} s^r_{44} s^u_{42})^n s^l_{21}
\nonumber \\
& = & s^r_{64} s^u_{42}
      ( 1 - s^l_{22} s^u_{24} s^r_{44} s^u_{42})^{-1}
      s^l_{21}
.\label{su61} \end{eqnarray}
Analogously, we can write $s^d_{61}$ as
\begin{equation}
s^d_{61} = s^r_{65} s^d_{53} 
( 1 - s^l_{33} s^d_{35} s^r_{55} s^d_{53})^{-1}
s^l_{31}
.\label{sd61} \end{equation}

From (\ref{propleft}), (\ref{propright}),
(\ref{su61}) and (\ref{sd61}) we obtain
\begin{equation}
s^d_{61} s^{u\dagger}_{61} = 0, \hspace{1cm} 
s^{d\dagger}_{61} s^u_{61} = 0
\label{cond2}
\end{equation}

The calculation of $s_{61}$ is simplified by the fact that
the conditions (\ref{condizioni}) are such that an electron passing
through $\Sigma^d$ cannot be scattered into $\Sigma^u$, and viceversa.
Therefore, we just have  
\begin{equation}
s_{61} = s^u_{61} + s^d_{61}
.
\label{sadd}
\end{equation}


The conductance of the whole structure at 0~K
is evaluated according to Landauer and B\"uttiker
\cite{landauer57,buettiker88,stonszaf88}:
\begin{equation}
G = \frac{2 e^2}{h} {\rm tr}\{ s^\dagger_{61} s_{61} \}
.\label{gtot}
\end{equation}
Let $G_{\rm up}$ and $G_{\rm down}$ be the conductances of the system 
with the conductor $\Sigma^d$ or $\Sigma^u$ removed, respectively:
we have
\begin{equation}
G_{\rm up} = \frac{2 e^2}{h} {\rm tr}\{ s^{u \dagger}_{61} s^u_{61} \}
, \hspace{1cm}
G_{\rm down} = \frac{2 e^2}{h} {\rm tr} \{s^{d \dagger}_{61} s^d_{61} \}
.\label{g2}
\end{equation}

From (\ref{cond2}) and (\ref{sadd}) we have
\begin{equation}
s^\dagger_{61} s_{61} =
(s^{u \dagger}_{61} + s^{d \dagger}_{61} )(s^{u}_{61} +
s^{d}_{61} ) 
= s^{u \dagger}_{61} s^u_{61} + s^{d \dagger}_{61} s^d_{61}
,\label{scat}
\end{equation}
which allows us to write, from (\ref{gtot}) and (\ref{g2}),
\begin{equation}
G = G_{\rm up} + G_{\rm down}
.\end{equation}

The additivity of conductances for parallel mesoscopic conductors
has been demonstrated. It implies also the additivity of thermal
noise current spectral densities, which are proportional to the
condunctance by means of a factor $4 k_B T$, where $k_B$ is
Boltzmann's constant and T is the absolute temperature.


The shot noise current spectral density in mesoscopic systems
can be expressed in terms of the scattering matrix as
derived by B\"uttiker \cite{buettiker90,buettiker92}:
\begin{eqnarray}
\langle ( \Delta I )^2 \rangle & = & 4|eV| \frac{e^2}{h} 
{\rm tr} \{ s^{\dagger}_{11} s_{11} s^{\dagger}_{61} s_{61} \}
\nonumber \\
& = & 
4|eV| \frac{e^2}{h} ( {\rm tr} \{ s^{\dagger}_{61} s_{61} \}
- {\rm tr}  \{ s^{\dagger}_{61} s_{61} s^{\dagger}_{61} s_{61} \} )
\label{noise}
\end{eqnarray}
where $V$ is the voltage applied between leads 1 and 6, and the
last equality comes from the unitarity of the matrix $S_{\rm tot}$.

The shot noise current spectral densities
$\langle (\Delta I_{\rm up})^2 \rangle$ and
$\langle (\Delta I_{\rm down})^2 \rangle$ of the system with $\Sigma^d$
or $\Sigma^u$ removed, respectively, are
\begin{eqnarray}
\langle (\Delta I_{\rm up})^2 \rangle = 
4|eV| \frac{e^2}{h} ( & & {\rm tr} \{ s^{u \dagger}_{61} s^u_{61} \}
\nonumber \\
& & 
- {\rm tr}  \{ s^{u \dagger}_{61} s^u_{61} s^{u \dagger}_{61} s^u_{61} \} )
\nonumber \\
\langle (\Delta I_{\rm down})^2 \rangle =
4|eV| \frac{e^2}{h} ( & & {\rm tr} \{ s^{d \dagger}_{61} s^d_{61} \}
\nonumber \\
& & 
- {\rm tr}  \{ s^{d \dagger}_{61} s^d_{61} s^{d \dagger}_{61} s^d_{61} \} )
.\label{noisepart}
\end{eqnarray}

From (\ref{cond2}) and (\ref{sadd}) we have
\begin{equation}
s^{\dagger}_{61} s_{61} s^{\dagger}_{61} s_{61}
=
s^{u \dagger}_{61} s^u_{61} s^{u \dagger}_{61} s^u_{61}
+
s^{d \dagger}_{61} s^d_{61} s^{d \dagger}_{61} s^d_{61} 
,\end{equation}
that, along with (\ref{scat}), (\ref{noise}) and (\ref{noisepart}),
allows us to write
\begin{equation}
\langle (\Delta I)^2 \rangle
= \langle (\Delta I_{\rm up})^2 \rangle 
+
\langle (\Delta I_{\rm down})^2 \rangle
,\end{equation}
i.e., shot noise current spectral densities add for mesoscopic conductors
connected in parallel by means of junctions satisfying (\ref{condizioni}).


The above demonstration has important consequences on the functionality
of devices based on the Aharonov-Bohm effect, even if only
one of the junction obeys to the conditions (\ref{condizioni}).
Without losing generality, let us suppose that $s^l_{32} = 0$ and
$s^l_{23} = 0$, which implies that (\ref{propleft}) holds true.

Since paths from lead 2 to 3 passing through the left junction are
not allowed, the matrix $s_{61}$ can be written as the sum of two terms
$s^{(2)}_{61}$ and $s^{(3)}_{61}$, which take into account only the
paths passing through $\Sigma^l$ from 1 to 2, and from 1 to 3,
respectively, and can be expressed in the form
\begin{equation}
s^{(2)}_{61} = s_{62} s^l_{21},
\hspace{1cm}
s^{(3)}_{61} = s_{63} s^l_{31}
,\end{equation}
where $s_{62}$ considers all the paths from lead 2 to 6, and $s_{63}$
all the paths from lead 3 to 6.
From (\ref{propleft}) we have 
\begin{equation}
s^{(2) \dagger}_{61} s^{(3)}_{61} = 0
\hspace{1cm}
s^{(3) \dagger}_{61} s^{(2)}_{61} = 0
\end{equation}
which allows us to write
\begin{eqnarray}
{\rm tr} \{ s^\dagger_{61} s_{61} \}
& = & 
{\rm tr} \{ ( s^{(2) \dagger}_{61} + s^{(3) \dagger}_{61} )
            ( s^{(2)}_{61} + s^{(3)}_{61} ) \} \nonumber \\
& = & 
{\rm tr} \{ s^{(2) \dagger}_{61} s^{(2)}_{61} \}
+
{\rm tr} \{ s^{(3) \dagger}_{61} s^{(3)}_{61} \}
.\label{finale}
\end{eqnarray}

Eq. (\ref{finale}) establishes that there is no interference between
paths included in $s^{(2)}_{61}$ and in $s^{(3)}_{61}$, i.e., the
associated conductances just add. In other words, devices based on
the modulation of conductance due to quantum interfence between
different paths cannot work if just one of the junction satisfies
the properties indicated in (\ref{condizioni}). This consideration
has to be taken into account, for example, when devices are designed
which exploit some sort of Aharonov-Bohm effect. 

For example, Aharonov-Bohm devices should have not even one junction
of the type shown in the inset of Fig. 2.: as can be seen the
term ${\rm tr}\{ s_{32}^\dagger s_{32} \}$, though not exactly zero,
is more than two orders of magnitude smaller than ${\rm tr}
\{ s_{12}^\dagger s_{12} \}$, which implies a vanishingly small
quantum interference. Devices with junctions similar to
that of Fig.~2 have been proposed in the literature \cite{datta90}.


We have shown that conductances, thermal and shot noise current 
spectral densities for parallel mesoscopic conductors add,
provided a few conditions are satisfied by the junction scattering
matrices. Our demonstration is trivially extended to the case of
more than two conductors in parallel. 

Furthermore, our considerations have important consequences on the
design of Aharonov-Bohm devices, and in particular on the shape 
of the junction connecting the different paths. Finally, 
we have shown an example of a junction warranting additivity of
conductance.


The present work has been supported by
the Ministry for University and Scientific and 
Technological Research and by the Italian National Research Council
(CNR).

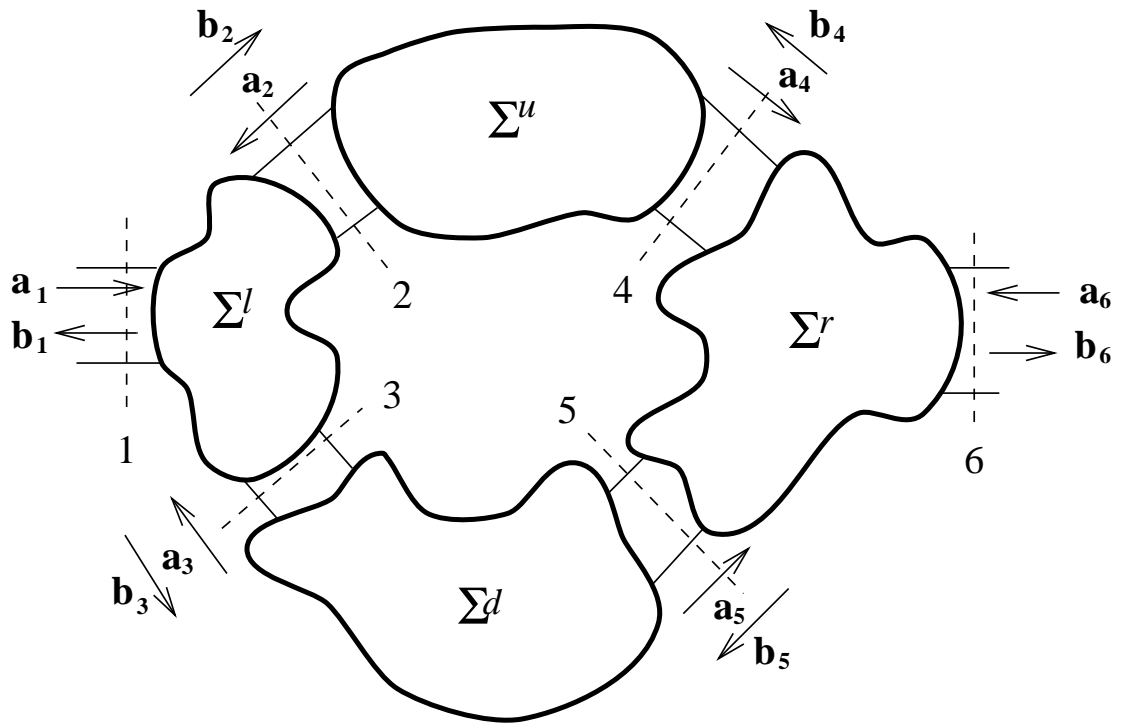
\begin{figure}
\caption{
The structure consists of 2
mesoscopic conductors $\Sigma^u$ and $\Sigma^d$ connected in parallel
by means of two junctions, $\Sigma^l$ and $\Sigma^r$. The junctions
are ballistic systems with three leads, one connected to $\Sigma^u$,
one to $\Sigma^d$, and the other used as an external lead of the 
whole structure. Phase coherence is mantained in the whole system.
${\bf a_i}$ and ${\bf b_i}$ ($i=1,\dots,6$)
are the column vectors whose $N_i$ elements are the amplitudes of the modes
in lead $i$ entering and exiting the adiacent junction, respectively.
}
\end{figure}

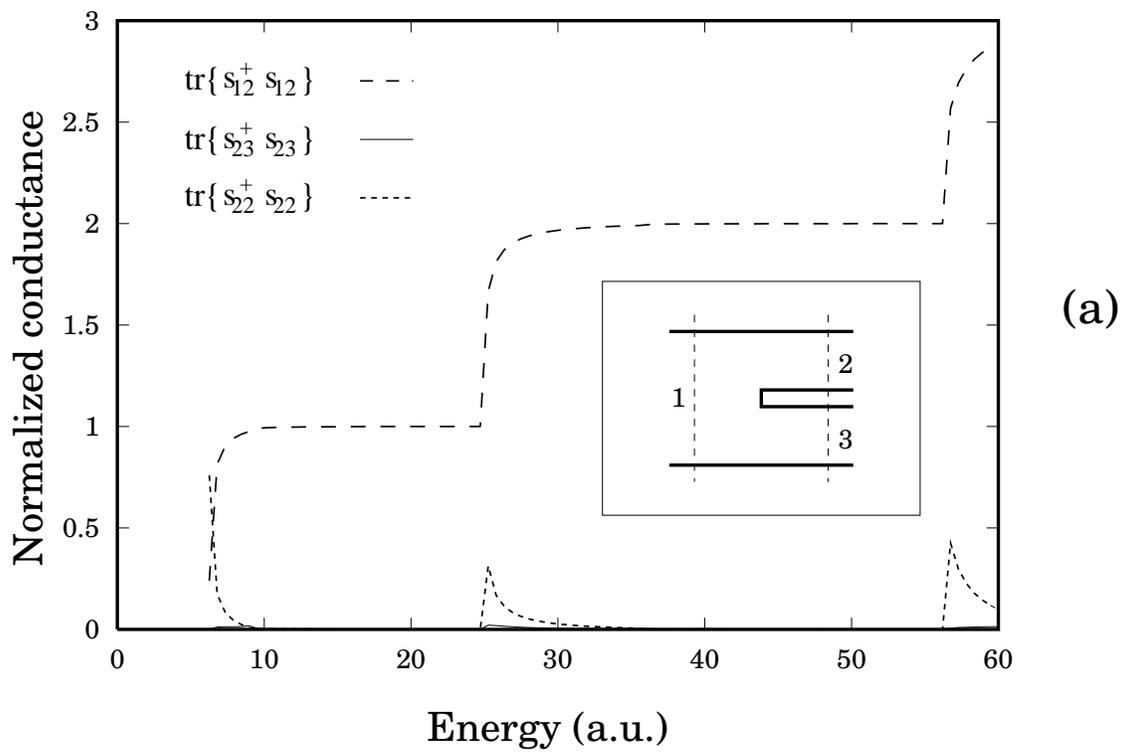
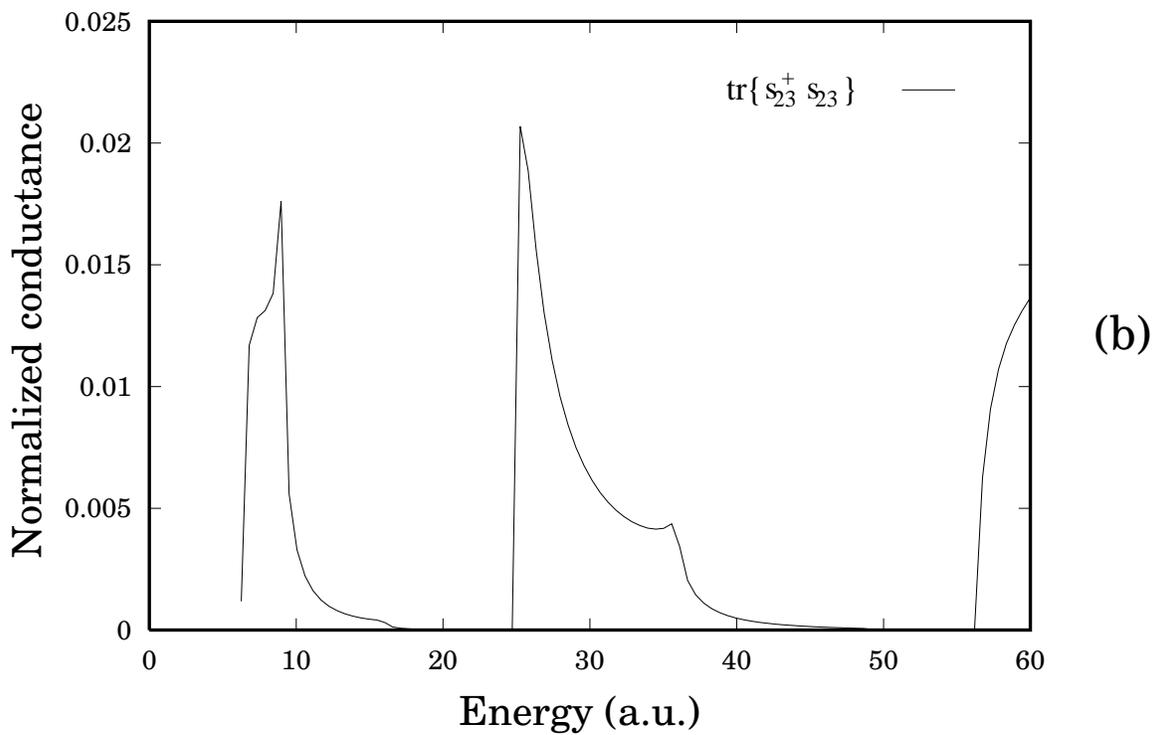
\begin{figure}
\caption{
The geometry of the junction is shown in the inset of Fig. (a).
In (a) ${\rm tr} \{s_{i2}^\dagger s_{i2} \}$, $i=1,2,3$ are
plotted as a function of the electron energy. In (b)
${\rm tr} \{ s_{32}^\dagger s_{32} \}$ is plotted on
a smaller scale. ${\rm tr} \{s_{12}^\dagger s_{12} \}$ has
the well known staircase behaviour while
${\rm tr} \{ s_{32}^\dagger s_{32} \}$ is at least two orders
of magnitude smaller.
}
\end{figure}

\end{document}